# Predicting Potential School Shooters from Social Media Posts

Alana Cedeno
Department of Computing Sciences
University of Hartford
West Hartford, USA
alana0114@gmail.com

Rachel Liang
Department of Computing Sciences
University of Hartford
West Hartford, USA
rachelrliang@gmail.com

Sheikh Rabiul Islam
Department of Computer Science
Rutgers University - Camden
Camden, USA
sheikh.islam@rutgers.edu

*Abstract*—The rate of terror attacks has surged over the past decade, resulting in the tragic and senseless loss or alteration of numerous lives. Offenders behind mass shootings, bombings, or other domestic terrorism incidents have historically exhibited warning signs on social media before carrying out actual incidents. However, due to inadequate and comprehensive police procedures, authorities and social media platforms are often unable to detect these early indicators of intent. To tackle this issue, we aim to create a multimodal model capable of predicting sentiments simultaneously from both images (i.e., social media photos) and text (i.e., social media posts), generating a unified prediction. The proposed method involves segregating the image and text components of an online post and utilizing a captioning model to generate sentences summarizing the image's contents. Subsequently, a sentiment analyzer evaluates this caption, or description, along with the original post's text to determine whether the post is positive (i.e., concerning) or negative (i.e., benign). This undertaking represents a significant step toward implementing the developed system in real-world scenarios.

## I. Introduction

Over the last few years, there has been an alarming increase in mass killings committed by individuals in the United States of America. Many of the perpetrators of these lone-actor attacks [1] utilize firearms to commit their horrible deeds. In 2022, at least 647 mass shootings [2] occurred in the United States, with four or more victims killed or injured. This represents a shocking 122.305% increase over the number of documented mass shootings in 2014 [3]. Since the beginning of 2023 until the time of writing this report, November 14, 2023, there have been a total of 602 mass shootings reported in the United States. Found in the Appendix, Fig. 6 depicts the total number of mass shootings in the United States from January 1, 2023 to November 14, 2023 [4].

Children are the group most at risk in this country, and they have been forever impacted by the bloodiest mass shootings in history. Compared to four years ago, the number of school shootings has nearly doubled [5]. With the median age of school shooters at 16 years old [6], and the prevalence of technology in today's world where many younger generations share their innermost thoughts online, we have never had greater access to the minds of mass killers. Lone-actor terrorists' social media posts have horrifyingly revealed the shooter's warped beliefs, obvious admiration of previous mass murderers, or even their future plans to carry out mass killings. Furthermore, the gunmen typically leave clues prior to the attack by sharing pictures of themselves holding and exhibiting their arsenal of weapons [7]. The possibility of attacks may be reduced if certain warning indicators can be identified [8]. This country has never before had access to such a valuable tool for stopping mass killings. However, it can be difficult to identify a possible lone-actor attacker due to the volume of social media posts that are uploaded every day.

We propose a system that combines an image classification model and a natural language processing model. This system will be able to fully analyze a social media post, including the textual content included in a typical online post as well as the image (seen on image-based social media platforms like Facebook or Instagram) and generated captions. A dataset with text as well as images was used to train each model in the integrated system, which allows social media posts to be evaluated as positive or negative. The curated dataset is available for public use and to support ongoing sentiment analysis research.

## II. Related Works

Although our project is focused on school shooters, with a unique dataset produced from actual online postings by the gunman prior to the mass murder, it is crucial to emphasize that the overall psychopathology of a school shooter has been documented in other violent criminals as well. Other murderers have praised the extreme ideologies and unusual conduct that our dataset of school shooters exhibited. Consequently, a wide range of additional lone-wolf attackers, including domestic terrorists and those who plot homicide or other violent crimes, may be covered by our system.

In the beginning, we decided to use natural language processing in our system to identify if a particular text was safe or cause for alarm. A substantial amount of research has been done about the sentiment analysis of a text. Among the studies we reviewed, Naive Bayes and Bidirectional Encoder Representations from Transformers (BERT) were investigated [9]-[10]. Furthermore, based on real data from the Twitter API, Go et al. created a dataset of over 1.6 million tweets to determine the sentiment of each tweet based on the emoticon included in the text. A tweet would be considered negative if there was a frowning face emoji, and positive if there was a happy face emoticon [9].

Chiorrini et al. investigated how BERT can be used for sentiment analysis and emotion recognition. They used a Twitter dataset with 1.6 million tweets for their sentiment analysis. Anger, happiness, fear, and sadness were some of the emotions that Chiorrini et al. included in their dataset of tweet emotion intensity. Following model training, it was discovered that the emotion recognition model had an accuracy of 90% while the sentiment analysis model had a 92% accuracy [10]. Instead of using the entire 1.6 million tweet dataset, our research will use BERT and train it on our text dataset that is specific to posts made by school shooters, as well as random sample tweets from the 1.6 million Twitter dataset.

The second aspect of our proposed system consists of an image classification model that can identify the context, subject, and action of each image. Deep learning models such as Visual Geometry Group (VGG), DenseNet, and ResNet have been used in several earlier studies on image sentiment analysis to assign a picture to a certain category [11].

Gajarla and Gupta researched how to categorize a picture as love, happiness, violence, fear, and sadness. Three pre-trained models were fine-tuned: VGG-ImageNet, VGG-Places205, and ResNet-50. These models were trained on a Flickr dataset that they created by using the API service and querying for images depending on the emotions that they had categorized. The ResNet-50 model achieved an accuracy of 73%, which was approximately 6% higher than the second most accurate model, VGG-ImageNet. However, they discovered that 80% of the images classified as happiness were face images, indicating that there is some bias in this category [11]. For our project, we considered a sentiment-based model for our system. However, we chose to create an image captioning model to obtain the context of each image, which performed better in discriminating between positive and negative with our own unique dataset based on common subjects discovered in prior school shooters' social media posts.

Combining two different types of models, one trained in text and the other in images, we created a 'multimodal' model. Multimodal learning, in which many modalities are used to process information, improves better understanding and analysis of data than unimodal learning in general. As humans, we experience our surroundings in a multimodal manner, using our various senses to gather more information and gain a better understanding of them. Machines, like humans, can evaluate and understand data from numerous media forms, such as images and text, by merging unimodal models that specialize in one type of media. This model combination enables more effective learning and improvement in data classification [12].

Numerous research have shown how the multimodal approach, as opposed to an unimodal, can analyze and produce better learning results. When compared to an unimodal strategy, it has been observed that a multimodal approach improves learning, reduces training time, and results in higher performance [13]. Our approach is a reflection of the fact that multimodal models typically result in better outcomes.

### III. METHODOLOGY

The proposed model, as shown in Fig. 1, will accept a social media post, which normally includes both text and an image. These posts will be separated into text and images, and the image will then be subject to an image captioning model. This image captioning model will generate a sentence summarizing the contents and context of the photo in order to understand the scene, as opposed to relying on object detection to highlight where an object is placed. This caption will then be joined with the text from the original post and placed through a sentiment analysis tool to determine if the post is positive (i.e., concerning) or negative (i.e., benign). The accuracy of this analysis is calculated by finding the average of how many times the model can correctly predict the sentiment of a social media post.

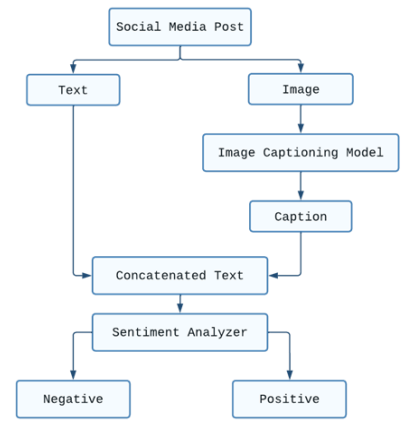

Fig. 1. Overall architecture of the proposed system.

### IV. DATA COLLECTION AND PROCESSING

We curated and used two media types (i.e., modalities) of datasets in our experiment.

#### A. Image Dataset

Websites like School Shooters.info [14] gather and archive all of the known social media posts made by school shooters into a single database, despite the fact that many of their social media accounts have been deleted or made private. We were able to examine similar visual elements in the gunmen's social media accounts by using these archived posts. Many of the perpetrators shared images of guns, gruesome pictures of dead animals, images of other weapons including knives, and images of gore/blood. Although we obtained the shooters' images to add to the image collection, there were insufficient images to reliably train an image captioning model. As a result, we decided to broaden our scope and look at mass shooters' social media posts because their visual content is similar. A social news website called Reddit offers a number of forums where people from different communities may express their interests, ask questions, and discuss anything of interest. A forum on Reddit dedicated to mass murderers discusses the phenomenon of mass murder, the perpetrators who commit them, their motivations, crimes, and the psychology of these individuals. Its purpose is to encourage discussion and an effort to comprehend human behavior [15].

To effectively train the model and to improve our image dataset, we applied augmentation techniques to artificially increase image size. This involved modifying the photos by adjusting them, thus creating new, altered versions of images sourced from the dataset. The aim was to enhance diversity, providing supplementary information for the machine learning process. These changes included changing an image's color, rotation, brightness, and contrast, as well as using techniques like flipping and cropping.

Taking the *positive class* into account, 77 images were categorized as school shootings, while 69 images were labeled as mass shootings. These images are in their original, non-augmented form, giving the positive class a total of 146 original images. We collected this dataset manually, sourcing as much publicly available information as possible.

Using the *augmentation* techniques stated above, we have 77 augmented images from the school shootings category and 69 augmented images from the mass shootings category.

This results in the addition of 146 augmented photos. 500 random images from the Flickr 8k dataset were included to enhance the dataset with a negative class representing non-threatening content [16]. This would provide greater context for the image captioning model to consider when analyzing an image.

To evaluate differently trained models, we divided our dataset into the following categories: non-threatening (negative class), mass shootings (positive class), school shootings (positive class), and two sets of *augmented* images from each of the positive classes. This allowed us to combine many categories and use them to train a captioning model to determine which combination produces the best results.

Five distinct human-made captions were developed for each photo in the dataset. The image and all five captions associated with it were used to train the model. Two people hand-labeled the unique images collected from real social media posts made by school shooters and mass shooters. Each sentence in the set of captions per image was unique, providing a range of possibilities for a model to describe an image. Fig. 2 shows an example of an image and the five captions that were written.

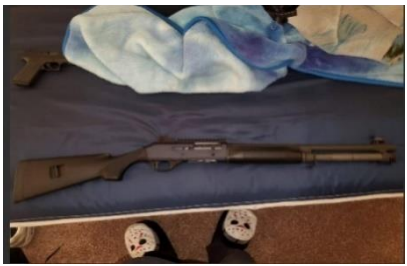

1) There is a person standing in front of two guns located over a bed .
2) A person stands next to a gun wearing slippers that have a white mask .
3) There is a individual standing in front of two guns wearing slippers that have the jason mask .
4) Two guns laying over a blue sheet .
5) Person stands looking over two guns located over a bed, one is cover by a blue blanket .

Fig. 2. Sample from image dataset shows image and custom captions associated with the image.

The captions that accompany the enhanced images were obtained by utilizing the captions from the original image dataset and augmenting them through backtranslation. This kind of augmentation is the process of re-translating text from the target language back to its source language. Given that the images were altered, back translation was used to generate artificial data from the original captions, resulting in an entirely artificial dataset containing image and text. The Helsinki-NLP models, notably the English-to-French and French-to-English models, were chosen for this project. These models are part of OPUS-MT, a project dedicated to the creation of free machine translation tools and resources. It employs Marian-NMT2 as its framework since it is a stable production-ready Neural Machine Translation (NMT) toolbox written in C++ with few dependencies and capable of efficient training and decoding. It also includes cutting-edge NMT architectures such as deep RNN and Transformers [17]. Over 1,000 pre-trained translation models are available for free download and use through the OPUS-MT project [18].

Using the original captions, the text was converted to a list by splitting after every new line and formatted into the target language to prepare the data for augmentation. Following that, the procedure of back translating the captions was straightforward. The first step was to translate the content from English into its temporary language, French. The temporary text was then translated back into its original language, English. Refer to Fig. 7 in the Appendix for an example of an augmented image along with five captions that were back-translated. After that, all the captions were pre-processed. This meant changing all captions to lowercase, removing digits and special characters, and removing additional spaces. Finally, the texts were tokenized to enable recognition by the model during training.

*B. Textual Dataset*

The text in the School Shooters.info archive was used to generate the positive class, text written by perpetrators, of our textual dataset. Words taken from social media like Instagram, YouTube, and Facebook postings were used in our dataset with the purpose of obtaining real data for the positive class. Notably, much of the textual content from these shooters shared similar philosophy across numerous accounts and people. Textual posts frequently included previous school shooters, hostile statements, threats of violence, and weaponry such as knives or guns. A total of 465 textual posts were extracted from the social media posts from school shooters and mass shooters, and added to our collection to be labeled as positive. We extracted a similar quantity of negatively tagged text from a pre-existing dataset of Twitter postings to populate the text dataset with benign text [19]. The textual database comprised a total of 328 online posts which were labeled as negative. Table 1 shows a sample of the dataset (before pre-processing). On the textual dataset, a similar pre-processing technique was used as on the image dataset captions. The text was stripped of usernames (@), hashtags (#), links, stop words, and other noise-related phrases.

TABLE I. EXAMPLE OF TEXTUAL DATASET, LABELED NEGATIVE (0) AND POSITIVE (1).

| Label | Text |
|---|---|
| 0 | listening to some music and just chilling....I'll probably regret not getting work done...but till then i'm just gonna kick back |
| 0 | i am working on my media room design and i love love love my client profile |
| 1 | more than anything I wish I could ve seen your faces and fought alongside |
| 1 | im going to be a professional school shooter |

V. EXPERIMENTATION

Several algorithms and techniques were tested. Each model's metrics would be analyzed and the most accurate model would be chosen for the entire system later on.

*A. Image Analysis: VGG-16*

For the purpose of captioning images, VGG-16 algorithm was chosen. VGG-16 is a Convolutional Neural Network (CNN) with 16 layers. The architecture of VGG-16 is made up of many uniformly positioned 3×3 filters [20]. As mentioned earlier, our unique dataset with handwritten captions for each photo served as the training set for the image captioning model. Contextual clues and other details about the image were given in each caption. This model was trained to be able to take an image, analyze it thoroughly, and generate a caption that describes the situations of the individuals in the picture.

As mentioned previously, we divided our dataset into the following categories: non-threatening (negative class), mass shootings (positive class), school shootings (positive class), and another two sets of *augmented* images from each of the positive classes, which allowed us the opportunity to combine many categories and use them to train a captioning model to determine which combination produces the best results. The combinations that we made are presented in Table 2.

TABLE II. COMBINATION OF CATEGORIES FOR TRAINING MODEL

| Combination | Number of images | Number of captions |
|---|---|---|
| Unedited mass shootings and non-threatening pictures | 569 | 2,845 |
| Unedited school shootings and non-threatening pictures | 577 | 2,885 |
| Combined unedited school shootings, unedited mass shootings, and non-threatening pictures | 646 | 3,230 |
| Augmented mass shootings and non-threatening pictures | 569 | 2,845 |
| Augmented school shootings and non-threatening pictures | 577 | 2,885 |
| Combined augmented school shootings, augmented mass shootings, and non-threatening pictures | 646 | 3,230 |
| Combined all unedited and augmented images from both school shootings and mass shootings, and non-threatening pictures | 792 | 3,960 |

Although each subset of the dataset has a different number of images and captions, the focus of the image captioning model is to find which one of these combinations can generate a caption that is more accurate in describing the image and is written like human language. An example of images that were evaluated after the captioning models were trained are shown in Fig. 3. Furthermore, the preprocessed unedited captions along with the captions generated through back-translation can be found in Table 3. These images display a comparison between the model's predicted captions and the actual captions. For additional information about another example, refer to Fig. 8 and Table 7 in the Appendix section.

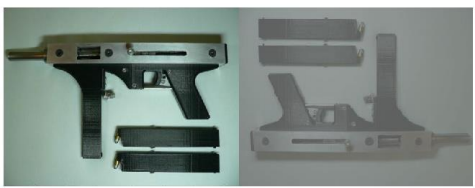

Fig. 3. Unedited image (left) taken from mass shooter's category and it's augmented image (right).

TABLE III. PREPROCESSED UNEDITED CAPTIONS AND BACK-TRANSLATED CAPTIONS FOR FIG.3

| Unedited captions | Back-translated captions |
|---|---|
| custom gun with gray and black parts | custom gun with grey and black pieces |
| there is custom gun with magazines | there's custom gun with magazines |
| custom made gun with two black magazines | gun made to measure with two black magazines |
| there can be seen custom gun with magazines places on flat surface | you can see custom gun with magazines placed on flat surface |
| gray and black custom gun with two magazines | gray and black custom gun with two magazines |

After training the captions on various combinations of categories to build variant models, the results were compared to determine which model most closely resembled human language. Table 4 provides predicted captions for each of the previously shown images in Fig. 3. For an additional example, refer to Table 8 in the Appendix section.

TABLE IV. PREDICTED CAPTION FOR FIG. 3 GENERATED BY IMAGE CAPTIONING MODELS

| Model combination | Predicted caption |
|---|---|
| Unedited mass shootings and non-threatening pictures | man is sitting on an of white |
| Combined unedited school shootings, unedited mass shootings, and non-threatening pictures | there is there is firearm to the camera |
| Augmented mass shootings and non-threatening pictures | a man is sitting on the beach |
| Combined augmented school shootings, augmented mass shootings, and non-threatening pictures | there is black standing next to be playing |
| Combined all unedited and augmented images from both school shootings and mass shooting, and non-threatening pictures | you can see custom weapon with magazines |

Looking at the predicted captions for both images, the model that combined all *unedited* and *augmented* images from both school shootings and mass shootings, as well as the non-threatening images, demonstrated an accurate and close to human language prediction. This means that the captioning model that was chosen for the proposed model was trained using the most images and captions, making it a total of 792 images and 3,960 captions.

Because we cannot evaluate a caption into direct binary decision like right or wrong, BiLingual Evaluation Understudy (BLEU) was used as a metric to evaluate the caption generated by the image captioning model. BLEU works on the idea that the closer a machine translation is to a professional human translation, the better. A BLEU score produces a number between 0 and 1, with the closer the score is to one, the closer the model is to a human translation [21]. It should be noted, however, that BLEU only examines direct word-to-word similarity and the degree to which word clusters in two sentences are identical. This suggests that even if the translation is correct, the model might have used a completely other term with the same meaning and still receive a low BLEU score. In practice, however, when two people come up with distinct sentence variations for a topic, they rarely accomplish a perfect word-to-word match. As a result, a BLEU score close to 1 is impractical in practice and should raise concerns that the model is overfitting, implying that the generated text is similar to at least one text from the training set. As a result, a model with a score of 0.6 or 0.7 is considered the most realistic [22].

Following training, it was discovered that the best image captioning model was with a combination of all *unedited* and *augmented* images from both school shootings and mass shootings, as well as the non-threatening pictures. It had a BLEU-1 score of 0.58 and a BLEU-2 score of 0.37. This shows that the image captioning model's image captions were satisfactory. Furthermore, it demonstrates that gradual increase of images can improve the model's performance during the training process for the purpose of obtaining a caption that is closer to human language.

## B. Text Analysis

Text analysis models includes:

*1) Naïve Bayes*

Naive Bayes is a probabilistic machine learning model for classification tasks such as sentiment analysis. This concept is founded on Thomas Bayes' Bayes Theorem that describes the relationship between the conditional probabilities of statistical quantities. However, Naive Bayes differs from this theorem in that it implies mutual independence of all attributes [23]. There are several Naive Bayes versions, including Gaussian, Complement, Multinomial, and Bernoulli. The variants that were evaluated for the sentiment analysis include Complement Naive Bayes (CNB), Multinomial Naive Bayes (MNB), and Bernoulli Naive Bayes (BNB). Multinomial Naive Bayes uses the Naive Bayes method with features drawn from a multinomial distribution. This distribution describes the probability of encountering counts in a variety of categories, such as word counts in a document. Complement Naive Bayes, on the other hand, was created to correct the assumptions made by the standard Multinomial Naive Bayes classifier due to skewed data [24]. Stated otherwise, this approach works well with datasets that are imbalanced, meaning that there are more samples in one class than in another. Finally, Bernoulli Naive Bayes is based on the binary concept of the simple occurrence of each phrase, resulting in a matrix of 0s and 1s. This implies that while it can determine whether the term is present or not, it is unable to calculate how frequently the term occurs. Furthermore, short texts perform better with this model than lengthy documents [25].

We experimented with several train/test splits and discovered that the 80/20 split on the dataset (634/159 split of text) was the most accurate ratio. With a test accuracy of 81.13%, Complement Naive Bayes proved to be the most accurate of the three Naive Bayes classifiers when applied to our dataset. For that reason, we reported CNB model's performance. The model's performance matrix is shown in Table 5. For the area under the curve refer to Fig. 9 located in the Appendix.

TABLE V. COMPLEMENT NAÏVE BAYES PERFORMANCE MATRIX

| CNB Results | Precision | Recall | F1-score | Support |
|---|---|---|---|---|
| Negative (0) | 0.83 | 0.68 | 0.75 | 65 |
| Positive (1) | 0.80 | 0.90 | 0.85 | 94 |
| Accuracy | | | 0.81 | 159 |
| Macro avg | 0.82 | 0.79 | 0.80 | 159 |
| Weighted avg | 0.81 | 0.81 | 0.81 | 159 |

*2) BERT*

To determine which sentiment analysis model was the most accurate, BERT, another popular natural language processing technique, was tested. We ran the model with every possible combination of BERT preprocessors and encoders, a total of 1,089 models. Given that there were 33 different pre-made BERT preprocessors and 33 different pre-made BERT encoders, we recorded the training and testing metrics of the final epoch to find the optimal combination of BERT preprocessors and encoders [26]. Through testing, it was discovered that the highest accuracy was also obtained with a training/testing split of 75/25 (595/198 split of text) for the text dataset.

After comparing all the BERT models, we selected the preprocessor bert_en_uncased_L-8_H-768_A-12 with the encoder small_bert/bert_en_uncased_L-8_H-768_A-12 as it had the highest accuracy of 88%. Table 6 shows the performance matrix of this model.

TABLE VI. BERT PERFORMANCE MATRIX

| BERT Results | Precision | Recall | F1-score | Support |
|---|---|---|---|---|
| 0 | 0.87 | 0.90 | 0.89 | 82 |
| 1 | 0.90 | 0.86 | 0.88 | 81 |
| Accuracy | | | 0.88 | 163 |
| Macro avg | 0.88 | 0.88 | 0.88 | 163 |
| Weighted avg | 0.88 | 0.88 | 0.88 | 163 |

The model's output is displayed in Fig. 4 for texts with positive, negative, and positive connotations, respectively. The closer the displayed number is to zero, the closer the text is to being categorized as positive.

```
I am going to shoot up the school
Positive: [0.3027264]
I had a lovely time today
Negative: [0.93201464]
I'm going to kill everyone here
Positive: [0.11110169]
```

Fig. 4. Sentiment analysis results from BERT for three different sentences.

## VI. RESULTING MULTIMODAL MODEL

We were able to predict the sentiment of a social media post that included both image and text by combining an image captioning model with a sentiment analysis trained on BERT. As a result, we created a multimodal model capable of determining if the entire post is positive or negative. Fig. 5 depicts the model being applied to a social media post from one of the accounts of a gunman from a bank heist [27]. The first sentence in the output is the text that is posted and the second sentence is the generated caption that describes the image.

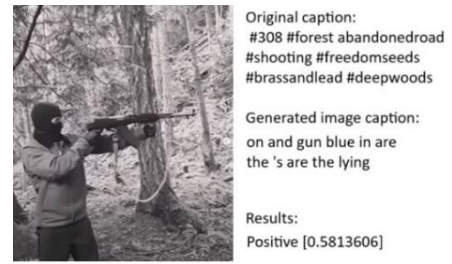

Fig. 5. Multimodal sentiment analysis result.

Looking at the generated caption shown in Fig. 5, it shows that the image captioning model correctly classified the social media post as positive (i.e., a concerning post). Although it can be treated as a working prototype with suboptimal performance, the model could be improved to enhance the language and readability of captions and overall accuracy of model with more training data and finetuning.

## VII. CONCLUSION AND FUTURE WORKS

We distinguished possible threats from social media posts (text and photos) using a multimodal method with acceptable accuracy and precision. To make the multimodal model more robust, the captioning dataset most likely would need to be expanded and can be an extension of this work. This involves improving the knife-to-gun ratio, as well as other subjects from the dataset's non-threatening side. While the image classifier is currently working well, there could be still room for improvement. Another possible improvement is training the image classifier to distinguish text from handwriting or typed texts within an image. In addition, future work would include expanding the text dataset to improve the BERT model's performance.

There is a chance that the model will demonstrate a higher false positive when evaluating postings that are not fully positive in the school shooting context, such as hunting, competitive shooting, and marketing. For a forecasting system in sensitive and high-stakes areas like predicting school shootings, a high rate of false positives is expected as a single successful prediction can save a lot of valuable lives.

The eventual goal for a real-world implementation and deployment of such a system would be to analyze real-time posts from multiple social media APIs such as Instagram and Twitter. Another priority to concentrate on is alerting the appropriate authorities when a post is flagged.

## VIII. OPEN SOURCING CODE AND DATASET

We made our developed dataset and code available to the public for further study in this area. The code and dataset can be accessed here: https://github.com/rrliang/Social-Media-Threat-Analysis

## ACKNOWLEDGMENT

Our thanks to Dr. Shubhra Kanti Karmaker (Santu), Assistant Professor, Auburn University for his valuable feedback on improving the paper from NLP perspective.

APPENDIX

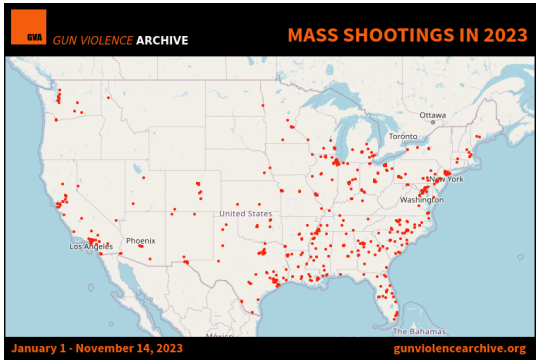

Fig. 6. Mass shootings in 2023 [4].

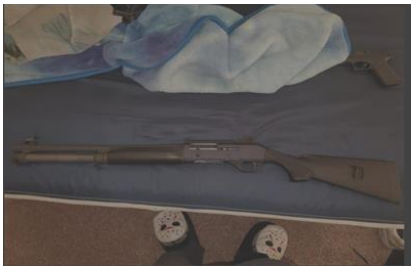

1) There is a person standing in front of two guns above a bed.
2) A person stands next to a gun with slippers that have a white mask.
3) There is an individual standing in front of two pistols carrying slippers that have the mask of jason.
4) Two guns placed on a blue leaf.
5) One person stands by looking over two guns above a bed, one is covered with a blue blanket .

Fig. 7. Sample from the image dataset shows an augmented image and captions associated with the image through back-translation.

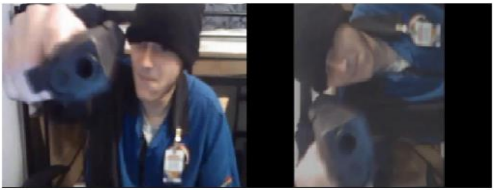

Fig. 8. Unedited image (left) taken from school shooter's category and it's augmented image (right).

TABLE VII. PREPROCESSED UNEDITED CAPTIONS AND BACK-TRANSLATED CAPTIONS FOR FIG.8

| Unedited captions | Back-translated captions |
|---|---|
| there is guy pointing gun at the camera in his bedroom | there's guy pointing gun at the camera in his room |
| there is man holding gun and pointing it at the camera | man holds weapon and the point is in the camera in front of bed |
| man wearing blue shirt is pointing gun | there's man holding gun and the point on the camera |
| man holds gun and points it at the camera in front of bed | man wearing blue shirt points weapon |
| there is guy aiming pistol at the camera | there's guy targeting gun on the camera |

TABLE VIII. PREDICTED CAPTION FOR FIG. 8 GENERATED BY IMAGE CAPTIONING MODELS

| Model combination | Predicted caption |
|---|---|
| Unedited school shootings and non-threatening pictures | the man is sitting in the air with catch ball |
| Combined unedited school shootings, unedited mass shootings, and non-threatening pictures | there is man in yellow shirt is looking at the camera |
| Augmented school shootings and non-threatening pictures | two men are sitting on the camera |
| Combined augmented school shootings, augmented mass shootings, and non-threatening pictures | two children are jumping in the background |
| Combined all unedited and augmented images from both school shootings and mass shooting, and non-threatening pictures | there's man pointing weapon on the camera |

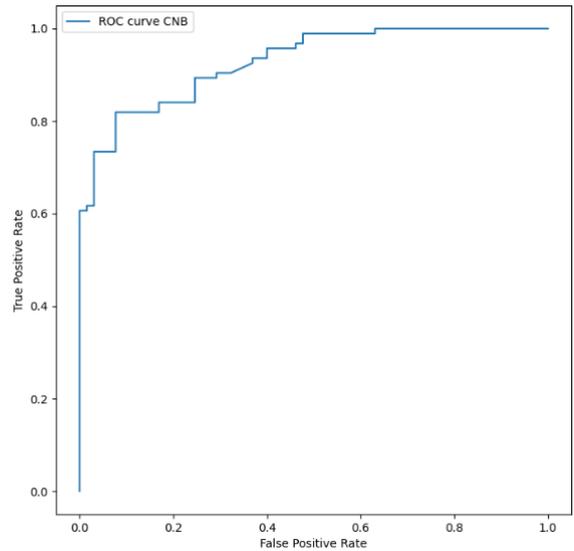

Fig. 9. Complement Naive Bayes Receiver Operating Characteristic curve (ROC).